\documentclass[aps,superscriptaddress,letterpaper]{revtex4}
\usepackage{amsmath,amssymb,graphicx,epsfig,color,bbm}
\usepackage[pass]{geometry}
\usepackage{pdfpages}

\def\nn{\nonumber}

\begin{document}



**Selective scattering between Floquet-Bloch and Volkov states in a topological insulator**

Fahad Mahmood[1], Ching-Kit Chan[1], Zhanybek Alpichshev[1], Dillon Gardner[1], Young Lee[1], Patrick A. Lee[1], Nuh Gedik[1†]

[1]Department of Physics, Massachusetts Institute of Technology, Cambridge, Massachusetts, 02139, USA

†To whom correspondence should be addressed (gedik@mit.edu)

The coherent optical manipulation of solids is emerging as a promising way to engineer novel quantum states of matter[1-5]. The strong time periodic potential of intense laser light can be used to generate hybrid photon-electron states. Interaction of light with Bloch states leads to Floquet-Bloch states which are essential in realizing new photo-induced quantum phases[6-8]. Similarly, dressing of free electron states near the surface of a solid generates Volkov states which are used to study non-linear optics in atoms and semiconductors[9]. The interaction of these two dynamic states with each other remains an open experimental problem. Here we use Time and Angle Resolved Photoemission Spectroscopy (Tr-ARPES) to selectively study the transition between these two states on the surface of the topological insulator $Bi_2Se_3$. We find that the coupling between the two strongly depends on the electron momentum, providing a route to enhance or inhibit it. Moreover, by controlling the light polarization we can negate Volkov states in order to generate pure Floquet-Bloch states. This work establishes a systematic path for the coherent manipulation of solids via light-matter interaction.

The manipulation of solids using ultrafast optical pulses has opened up a new paradigm in condensed matter physics by allowing the study of emergent physical properties that are otherwise inaccessible in equilibrium[1,2,10]. An important example are Floquet-Bloch states[11]

1

which emerge in solids due to a coherent interaction between Bloch states inside the solid and a periodic driving potential. This is a consequence of the Floquet theorem[12] which states that a Hamiltonian periodic in time with period $T$ has eigenstates that are evenly spaced by the drive energy ($2\pi\hbar/T$). Floquet-Bloch states have generated a lot of interest recently both for realizing exotic states of matter such as a Floquet Chern insulator[7] as well as understanding non-equilibrium periodic thermodynamics[13,14]. Experimental observation of these states requires the measurement of the transient electronic band structure of a crystal as it is perturbed by light. As has recently been demonstrated[15] in the topological insulator $Bi_2Se_3$, Time-and-Angle Resolved Photoemission spectroscopy (Tr-ARPES) is a key tool that can achieve this. Characteristic signatures of Floquet-Bloch states in the Tr-ARPES spectra include replicas of the original band structure that are separated by the driving photon energy[15].

In addition to Floquet-Bloch states, light can also generate other coherent phenomena in solids[16-18]. In particular, it can dress free electron states near the surface of a solid (Fig. 1a) since the surface can provide the momentum conservation necessary for a photon to interact with a free electron. This dressing was first observed in time resolved photoemission experiments[17] and has subsequently been referred to as Laser Assisted Photoemission (LAPE). LAPE is typically understood[19-21] by invoking the Volkov solution which is an exact solution of the time dependent Schrodinger equation for a free electron interacting with a plane electromagnetic wave[22]. LAPE bands can thus be thought of as Volkov states in vacuum that electrons can transition into from initial Bloch states inside the solid. In a Tr-ARPES experiment, the final state of photoemission is typically free electron-like and dressing of these final states generates Volkov states that, similar to Floquet-Bloch states, appear in the spectra as band replicas separated by the driving

2

photon energy. In this work we will refer to the dressing of initial states as Floquet-Bloch states and the dressing of the final states as Volkov states.

Both these dressed states cause band replicas in the Tr-ARPES spectra which appear at the same energy and momentum regardless of whether they originate from Floquet-Bloch or Volkov states, making it difficult to distinguish them. Moreover, due to the coherent nature of both processes, electrons can scatter directly from Floquet-Bloch states into Volkov states[23] (Fig. 1a). In order to study the various exotic affects predicted by Floquet theory on solid-state systems, it is important to experimentally characterize and separate out Floquet-Bloch and Volkov states in a controlled way. Furthermore, the interaction between Volkov and Floquet-Bloch states can provide novel insights in using semi-conductors for non-linear optics[9].

In this Letter, we use Tr-ARPES on $Bi_2Se_3$ with mid-IR excitation pulses to selectively study transitions between Floquet-Bloch and Volkov states. We find that interference between Floquet-Bloch and Volkov states must be taken into account to explain the intensity and the angular dependence of the dressed states in the Tr-ARPES spectra. Moreover, by controlling the polarization of the dressing field we can enhance or inhibit this interference. We also find that the observed hybridization between different dressed sidebands is independent of Volkov states and thus is a key signature of Floquet-Bloch states emerging in a solid.

Tr-ARPES measurements were performed using mid-Ir 160 meV pulses as the pump and 6.3 eV pulses as the probe. A time-of-flight analyzer is used to simultaneously acquire the complete transient band structure of $Bi_2Se_3$ without rotating the sample or the detector[24]. The mid-IR pump beam was incident on the sample at an angle of ~45° and its polarization was set to either P or S with respect to the incident plane (Fig. 1b). P-polarized pump includes an out-of-plane electric

field component whereas S-polarized pump is purely in plane. Figure 1c shows the Tr-ARPES spectra using P-polarized pump on $Bi_2Se_3$ at various delay times between the pump and probe. Replicas of the original Dirac cone appear when the pump and probe pulse overlap in time. These replicas are electron states dressed by the intense pump pulse. The intensity of these sidebands is maximized at $t = 0$ ps which refers to the maximum E-field of the pump beam coinciding with the maximum E-field of the probe. Once the dressing field of the pump pulse disappears ($t > 500$ fs), the sidebands disappear leaving a heated Dirac cone. The dynamics of this non-equilibrium heated distribution of electrons has been discussed in a number of Tr-ARPES experiments[25-28]. Here we will focus on the Tr-ARPES spectra taken at $t = 0$ ps to ascertain the relative contribution of Floquet-Bloch and Volkov states.

In order to disentangle the two, we study the Tr-ARPES spectra at $t = 0$ ps along various directions of the electron momentum. Figure 2a and 2b show the spectra along the $k_x$ and the $k_y$ directions respectively, taken with linear P-polarized pump with an in-plane electric field component along $k_x$ (Fig. 1b). Two observations are apparent: (1) avoided crossing gaps along $k_y$ (Fig. 2b, red arrows) but not along $k_x$ and (2) asymmetry in the intensity of Floquet sidebands about $k_x = 0$. The first observation is consistent with Floquet-Bloch theory on Dirac systems[29-32]. Since the pump E-field is along the x-direction, the perturbing Hamiltonian commutes with the Dirac Hamiltonian corresponding to electrons with momentum along $k_x$. This leads to a trivial crossing between sidebands along $k_x$ which thus remains gapless. However, along $k_y$, the direction perpendicular to the E-field, avoided crossing gaps open up. The gap ($2\Delta$) at the crossing between the zeroth and first order sideband is predicted[32] to scale linearly with the electric field amplitude ($E_0$) and thus $2\Delta \propto \sqrt{P}$, where $P$ is the applied average pump power. By plotting the measured value of the gap as a function of the pump power on a log-log plot (Fig.

4

2c), we find that 2Δ indeed scales as the square root of the pump power. This observation unequivocally establishes the transient generation of Floquet-Bloch states.

The second observation of asymmetry in the intensity of the sidebands allows us to establish scattering between Floquet-Bloch and Volkov states. As seen in Fig. 2a, the first order sideband ($n1$) is not an exact replica of the original band ($n0$). Rather, the replication of the Dirac cone is asymmetric between the $+k_x$ and $-k_x$ directions. It is important to distinguish this from the asymmetry in the intensity of the Dirac cone that arises in the unperturbed ARPES spectra. Due to the coupling of the photo-emitting 6 eV probe beam to the spin texture of the Dirac cone, there is a natural asymmetry between the $+k_x$ and $-k_x$ direction since the incident plane of the photo-emitting probe is along the $k_x$ direction. This matrix element (spin-probe) effect has been well understood in other ARPES measurements on similar systems[24]. Here, we will study the additional asymmetry that is present in the replica of the original Dirac cone. This asymmetry is more evident in constant energy cuts separated by the driving photon energy (Fig.3a, b). In order to minimize the effects of spin-texture as well as detector non-linearities, we divide these constant energy cuts ($I_1/I_0$) and plot the result in Fig.3c. If the $n1$ sideband were an exact replica of the $n0$ sideband, then $I_1/I_0$ would be constant as a function of the electron momentum. However, as can be seen in Fig.3c and in Fig.3d, $I_1/I_0$ is stronger along the $-k_x$ direction than along the $+k_x$ direction indicating that the dressed bands strongly depend on the direction of the electron momentum.

To explain this, we model our Tr-ARPES spectra by including the effects of both Floquet-Bloch and Volkov states. We start with the Dirac Hamiltonian describing the surface states of a topological insulator (supplementary). The mid-IR laser pump is introduced through the Peierl's

substitution i.e. $v_f \vec{k} \to v_f \vec{k} + e v_f \vec{A}$, where $\vec{A}$ is the vector potential of the pump light and $v_f$ is the Fermi-velocity. The dimensionless parameter $\beta = e v_f A / \omega$ characterizes the strength of the Floquet interaction, where $\omega$ is the frequency of the mid-IR laser pump. The resulting Tr-ARPES intensity can be obtained[33] by using the non-equilibrium two time correlation function of the driven electrons (supplementary). Without including the effect of Volkov states (LAPE), this results in the following expression for the photo-emitted intensity for the case of electron momentum along the linearly polarized pump direction, (i.e. along $k_x$):

$$I(k_x, E) \propto \sum_n \{\delta_{E, \hbar v_f k_x + n\hbar\omega} + \delta_{E, -\hbar v_f k_x + n\hbar\omega}\} J_n(\beta)^2 \qquad \text{eq.1}$$

Therefore, the *n*-th Floquet-Bloch sideband has an intensity $\sim J_n(\beta)^2$ and is symmetric for $\pm k_x$. The situation becomes different when the effect of Volkov states is included. The corresponding Hamiltonian is $H_{LAPE} = \hbar e \vec{v}_0 \cdot \vec{A}$, where $\vec{v}_0$ is the free photoelectron velocity. The dimensionless parameter $\alpha = e v_0 A / \omega$ characterizes the interaction strength between light and the final states of photoemission. The photo-emitted intensity (along $k_x$) now becomes (supplementary):

$$I(k_x, E) \propto \sum_n [\delta_{E, \hbar v_f k_x + n\hbar\omega} J_n(\beta - \alpha)^2 + \delta_{E, -\hbar v_f k_x + n\hbar\omega} J_n(\beta + \alpha)^2] \qquad \text{eq.2}$$

The dependence on both $\alpha$ and $\beta$ is due to the interference between Floquet-Bloch and Volkov states. The observed *n*-th order sideband is now a combination of different Fourier pairs of Floquet-Bloch ($n_F$) and Volkov ($n_V$) modes such that $n_F + n_V = n$ (Fig. 1a). In order to explain the data fully, we have also included the spin-probe effect that describes the coupling of the photo-emitting probe to the spin texture of the Dirac cone (supplementary).

Figure 3e shows the results of this calculation for three different cases: (i) $\alpha = 0$ & $\beta = 0.5$ (Floquet only), (ii) $\alpha = 1.38$ & $\beta = 0$ (Volkov only) and (iii) $\alpha = 1.38$ & $\beta = 0.5$ (Floquet & Volkov). The non-zero values used for $\alpha$ and $\beta$ agree quite well with the measured experimental parameters ($v_f$, $v_0$, A and $\omega$) of the setup (supplementary). In case i (red trace) electrons scatter from dressed states in the solid (Floquet-Bloch) into unperturbed free electron states. The two-fold rotational symmetry is understood by noting that the electrons scatter preferentially when their momentum is along the direction of the light polarization (in this case along $k_x$). Case ii (green trace) refers to the situation when only the final states are dressed (Volkov). Here it is important to note that in the photoemission process only the in-plane momentum is conserved whereas the electrons acquire a large out-of-plane momentum ($k_z$) due to the excess photon energy. Since the pump pulse is P-polarized, the electric field in the z-direction strongly couples to free electron states with a large $v_z$ leading to a dressing of these final states that predominately depends on the out-of-plane momentum. The intensity of the sidebands is thus isotropic as a function of in-plane momentum (green trace).

Case iii (blue trace) includes the dressing of both the initial and final states and, as can be seen, this trace closely matches the observed angular dependence in the intensity of the first order sideband (Fig. 3d). The calculation also captures the strong asymmetry in $I_1/I_0$ between $+k_x$ and $-k_x$, which would not be present for pure Floquet-Bloch (case i) or pure Volkov states (case ii). This result not only implies the presence of both Floquet-Bloch and Volkov states but also points to selective transitions between the two. For example, as the electron momentum is varied between $-k_x$ and $+k_x$, there is an increase and then decrease in the scattering intensity as recorded by Tr-ARPES. As discussed above, eq. 2 implies that for electron momentum along the light polarization direction (i.e. along $k_x$), the photo-emitted intensity can be written as $\propto J_n(\beta \mp \alpha)^2$

7

for $\pm k_x$. Thus, by varying the electron momentum, we can control the scattering between Floquet-Bloch and Volkov states. These selective transitions are a direct consequence of Volkov states being generated primarily due to the out-of-plane E-field for P-polarized pump.

The aforementioned result suggests a way to reduce the effect of Volkov states: eliminating the out-of-plane electric field. This can be achieved by perturbing the system with S-polarized light instead. Figure 4a and 4b show the Tr-ARPES spectra at $t = 0$ along the $k_x$ and $k_y$ directions respectively, taken with S-polarized pump oriented along the $k_y$ direction. Similar to the P-polarized pump, avoided crossing gaps are apparent (red arrows, Fig. 4a). However, the gaps are now observed along the $k_x$ direction but not along the $k_y$ direction. This is again consistent with Floquet-Bloch theory on Dirac electrons since the avoided crossing gaps are along the momentum direction ($k_x$) perpendicular to the direction of the E-field ($k_y$). We can also notice a significant decrease in the intensity of the sidebands despite using a similar intensity for the pump. We again take the ratio of the intensities of the first and zero order sidebands ($I_1/I_0$) and plot it as a function of the electron momentum direction (Fig. 4c). This ratio is almost 10 times less than what is observed for the P-polarized pump. We attribute this to a minimization of Volkov states which is confirmed by numerically calculating the sideband intensities case for the presence of Floquet-Bloch states only ($\alpha = 0$ & $\beta = 0.5$) using S-polarized pump. As can be seen (Fig. 4d), the calculation agrees quite well with the observed angular dependence of $I_1/I_0$. Thus, perturbing the system with S-polarized mid-IR pump results in the generation and observation of pure Floquet-Bloch states. Moreover, by controlling the light polarization, we can enhance or completely inhibit the transition between Floquet-Bloch and Volkov states.

In conclusion, we have used Tr-ARPES to directly observe selective transitions between Floquet-Bloch and Volkov states. These transitions strongly dependent on the initial electron

momentum as well as the polarization of the perturbing light pulse. The effects of Volkov states can be negated by using S-polarized pump which results in pure Floquet-Bloch states. Our work forges a systematic path to manipulate electronic states in a solid in order to realize exotic light-induced quantum states of matter in a variety of solid-state systems.

## Acknowledgements

## References:

1   Hu, W. *et al.* Optically enhanced coherent transport in YBa2Cu3O6.5 by ultrafast redistribution of interlayer coupling. *Nat Mater* **13**, 705-711, doi:10.1038/nmat3963 (2014).

2   Matsunaga, R. *et al.* Light-induced collective pseudospin precession resonating with Higgs mode in a superconductor. *Science* **345**, 1145-1149, doi:10.1126/science.1254697 (2014).

3   Kirilyuk, A., Kimel, A. V. & Rasing, T. Ultrafast optical manipulation of magnetic order. *Reviews of Modern Physics* **82**, 2731-2784 (2010).

4   Caviglia, A. D. *et al.* Ultrafast Strain Engineering in Complex Oxide Heterostructures. *Physical Review Letters* **108**, 136801 (2012).

5   Mentink, J. H., Balzer, K. & Eckstein, M. Ultrafast and reversible control of the exchange interaction in Mott insulators. *Nat Commun* **6**, doi:10.1038/ncomms7708 (2015).

6   Lindner, N. H., Refael, G. & Galitski, V. Floquet topological insulator in semiconductor quantum wells. *Nat Phys* **7**, 490-495 (2011).

7   Grushin, A. G., Gómez-León, Á. & Neupert, T. Floquet Fractional Chern Insulators. *Physical Review Letters* **112**, 156801 (2014).

8       Kitagawa, T., Oka, T., Brataas, A., Fu, L. & Demler, E. Transport properties of nonequilibrium systems under the application of light: Photoinduced quantum Hall insulators without Landau levels. *Physical Review B* **84**, 235108 (2011).

9       Wegener, M. *Extreme non-linear optics*. (Springer, 2005).

10      Stojchevska, L. *et al.* Ultrafast Switching to a Stable Hidden Quantum State in an Electronic Crystal. *Science* **344**, 177-180, doi:10.1126/science.1241591 (2014).

11      Faisal, F. H. M. & Kamiński, J. Z. Floquet-Bloch theory of high-harmonic generation in periodic structures. *Physical Review A* **56**, 748-762 (1997).

12      Galitski, V.M., Goreslavskii, S.P., Elesin, V.F. Electric and Magnetic Properties of a Semiconductor in the Field of a Strong Electromagnetic Wave. *JETP* **30** (1970).

13      Kohn, W. Periodic Thermodynamics. *Journal of Statistical Physics* **103**, 417-423, doi:10.1023/A:1010327828445 (2001).

14      Eisert, J., Friesdorf, M. & Gogolin, C. Quantum many-body systems out of equilibrium. *Nat Phys* **11**, 124-130, doi:10.1038/nphys3215 (2015).

15      Wang, Y. H., Steinberg, H., Jarillo-Herrero, P. & Gedik, N. Observation of Floquet-Bloch States on the Surface of a Topological Insulator. *Science* **342**, 453-457, doi:10.1126/science.1239834 (2013).

16      Schmitt, F. *et al.* Transient Electronic Structure and Melting of a Charge Density Wave in TbTe3. *Science* **321**, 1649-1652, doi:10.1126/science.1160778 (2008).

17      Saathoff, G., Miaja-Avila, L., Aeschlimann, M., Murnane, M. M. & Kapteyn, H. C. Laser-assisted photoemission from surfaces. *Physical Review A* **77**, 022903 (2008).

18      Miaja-Avila, L. *et al.* Ultrafast studies of electronic processes at surfaces using the laser-assisted photoelectric effect with long-wavelength dressing light. *Physical Review A* **79**, 030901 (2009).

19      Glover, T. E., Schoenlein, R. W., Chin, A. H. & Shank, C. V. Observation of Laser Assisted Photoelectric Effect and Femtosecond High Order Harmonic Radiation. *Physical Review Letters* **76**, 2468-2471 (1996).

20      Madsen, L. B. Strong-field approximation in laser-assisted dynamics. *American Journal of Physics* **73**, 57-62, http://dx.doi.org/10.1119/1.1796791 (2005).

21      Baggesen, J. C. & Madsen, L. B. Theory for time-resolved measurements of laser-induced electron emission from metal surfaces. *Physical Review A* **78**, 032903 (2008).

22  C. J. Joachain, N. J. K., R. M. Potvliege. *Atoms in Intense Laser Fields*. (Cambrige University Press, 2014).

23  Park, S. T. Interference in Floquet-Volkov transitions. *Physical Review A* **90**, 013420 (2014).

24  Wang, Y. H. *et al.* Observation of a Warped Helical Spin Texture in $Bi_2Se_3$ from Circular Dichroism Angle-Resolved Photoemission Spectroscopy. *Physical Review Letters* **107**, 207602 (2011).

25  Wang, Y. H. *et al.* Measurement of Intrinsic Dirac Fermion Cooling on the Surface of the Topological Insulator $Bi_2Se_3$ Using Time-Resolved and Angle-Resolved Photoemission Spectroscopy. *Physical Review Letters* **109**, 127401 (2012).

26  Sobota, J. A. *et al.* Ultrafast Optical Excitation of a Persistent Surface-State Population in the Topological Insulator $Bi_2Se_3$. *Physical Review Letters* **108**, 117403 (2012).

27  Hajlaoui, M. *et al.* Ultrafast Surface Carrier Dynamics in the Topological Insulator $Bi_2Te_3$. *Nano Letters* **12**, 3532-3536, doi:10.1021/nl301035x (2012).

28  Hajlaoui, M. *et al.* Tuning a Schottky barrier in a photoexcited topological insulator with transient Dirac cone electron-hole asymmetry. *Nat Commun* **5**, doi:10.1038/ncomms4003 (2014).

29  Syzranov, S. V., Fistul, M. V. & Efetov, K. B. Effect of radiation on transport in graphene. *Physical Review B* **78**, 045407 (2008).

30  Oka, T. & Aoki, H. Photovoltaic Hall effect in graphene. *Physical Review B* **79**, 081406 (2009).

31  Zhou, Y. & Wu, M. W. Optical response of graphene under intense terahertz fields. *Physical Review B* **83**, 245436 (2011).

32  Fregoso, B. M., Wang, Y. H., Gedik, N. & Galitski, V. Driven electronic states at the surface of a topological insulator. *Physical Review B* **88**, 155129 (2013).

33  Freericks, J. K., Krishnamurthy, H. R. & Pruschke, T. Theoretical Description of Time-Resolved Photoemission Spectroscopy: Application to Pump-Probe Experiments. *Physical Review Letters* **102**, 136401 (2009).

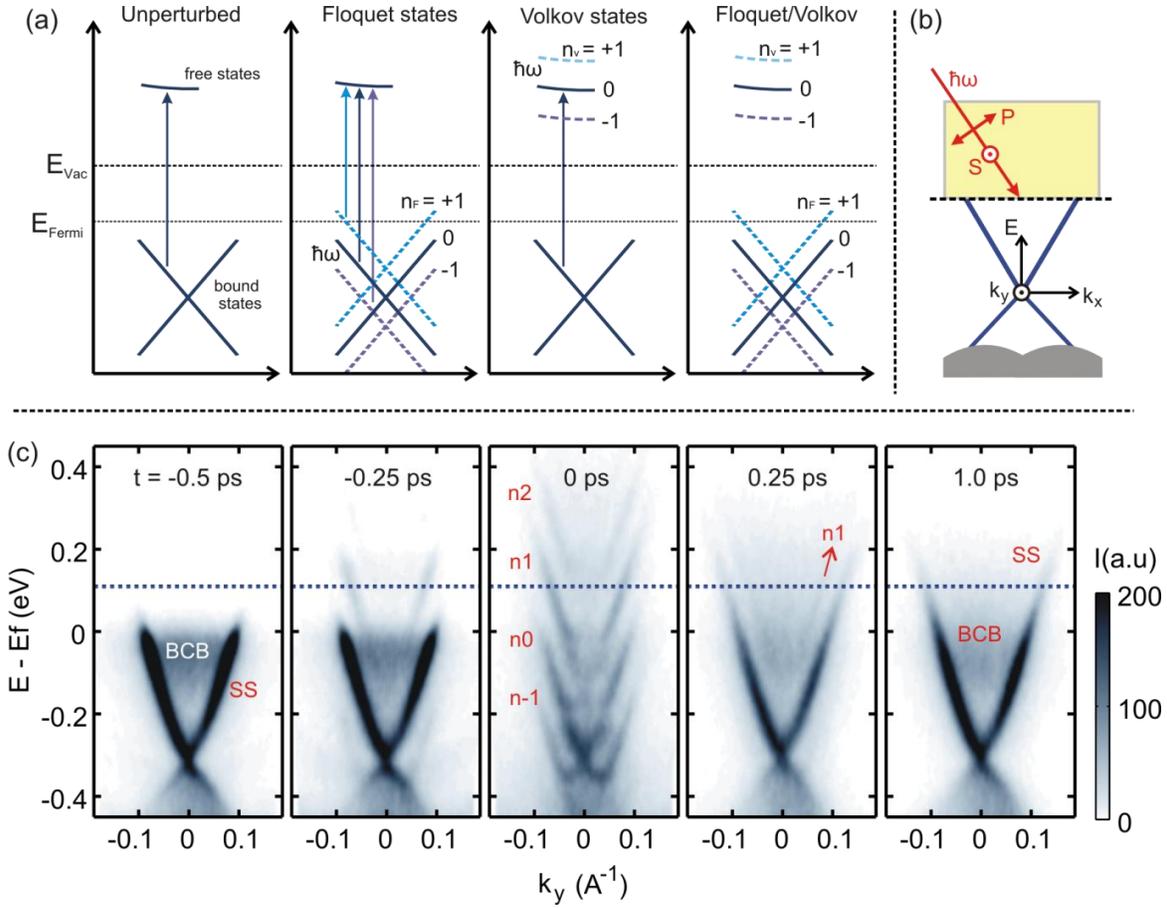

**Figure 1.** Dressed electron states in the Tr-ARPES spectra of a topological insulator and experimental geometry. **(a)** This schematic illustrates the various transitions in a Tr-ARPES experiment. In the unperturbed case or before time zero, electrons transition from bound states to free electron-like states. Dressing of the bound states results in Floquet states ($n_F$) separated by the drive photon energy $\hbar\omega$ whereas dressing of the free electron-like states results in Volkov states ($n_V$). If both the initial and final states are dressed (last panel), the $n^{th}$ order sideband in the Tr-ARPES spectra is given by all transitions such that $n = n_F + n_V$. **(b)** Experimental geometry of the Tr-ARPES setup. The pump light is incident onto the sample at an angle of ~45°. The pump is linearly polarized with P-polarization having an out-of-plane component and an in-plane component along the $k_x$ direction whereas S-polarization is purely in-plane along the $k_y$ direction. **(c)** Tr-ARPES (E-Ef vs $k_y$) spectra on $Bi_2Se_3$ using P-polarized pump at various delay times between the pump and the probe. BCB refers to the bulk conduction band while SS refers to the topological surface state. The $n^{th}$ order sidebands are indicated in the spectra at $t = 0$.

12

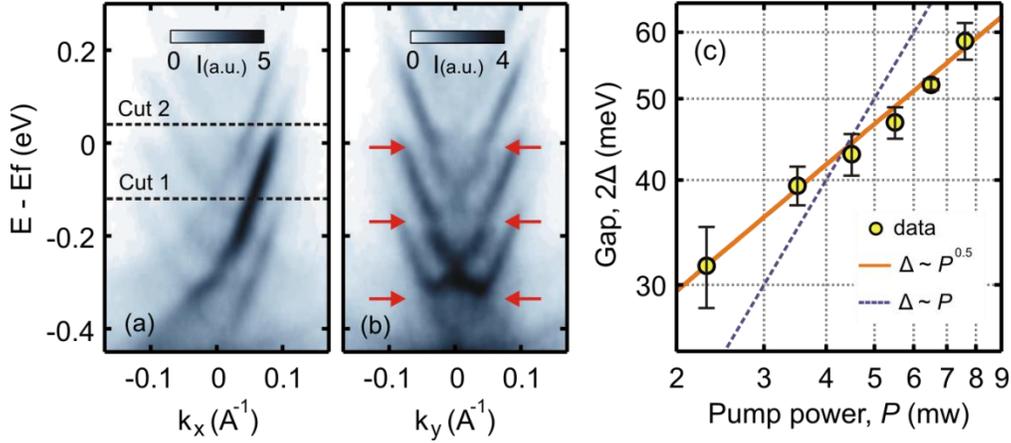

**Figure 2.** Tr-ARPES spectra of $Bi_2Se_3$ at $t = 0$ for P-polarized pump. **(a)** Energy (E) relative to the Fermi level (Ef) vs momentum along the $k_x$ direction and **(b)** along the $k_y$ direction. Red arrows indicate the avoided crossing gaps. **(c)** Avoided crossing gap ($2\Delta$) as a function of incident pump power (P) on a log-log plot. The gap at each pump power is obtained by fitting the energy distribution curves (EDCs) in the ARPES spectra with a pair of Lorentzians (supplementary text). Error bars represent the 95% confidence interval (2 s.d.) in extracting the gap from the fitting parameters. Power laws ($\Delta \sim P^\eta$) with $\eta = 0.5$ (orange trace) and $\eta = 1$ (blue trace) are plotted as well to determine the analytical behavior of $2\Delta$ with P.

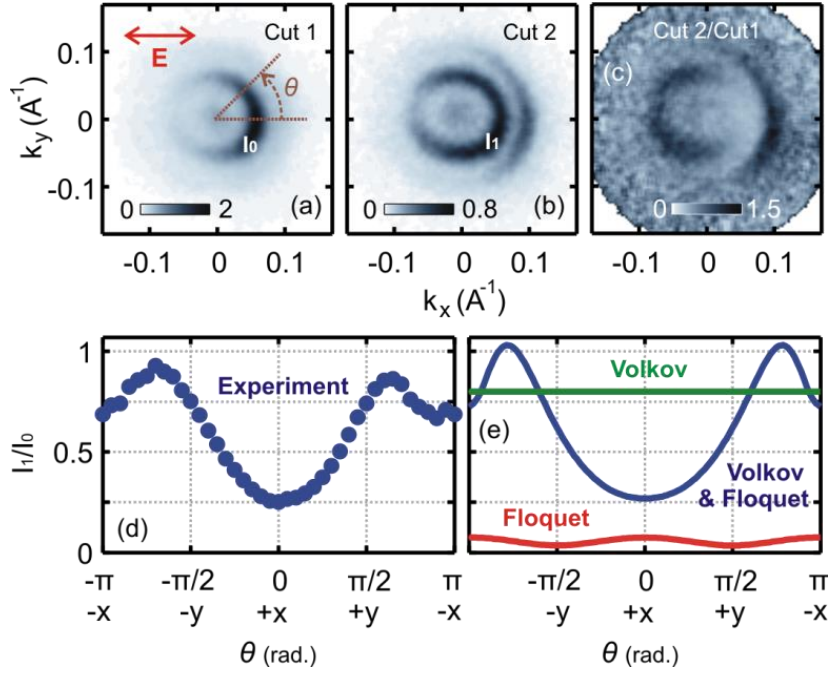

**Figure 3.** Asymmetry in the Tr-ARPES spectra. **(a)** Constant energy cut at E - Ef = -0.12 eV, i.e. along dashed line 'Cut 1' in Fig. 2a. The electric field ($\vec{E}$) is along the $k_x$ direction (red arrow). $I_0$ indicates the surface state contour for the zeroth order band i.e. the original Dirac cone. **(b)** Constant energy cut at E – Ef = 0.04 eV, i.e. along dashed line 'Cut 2' in Fig. 2a. The two constant energy cuts are separated in energy by the driving pump energy of 160 meV. $I_1$ indicates the surface state contour for the first order side band. **(c)** The constant energy cut in (b) is divided by the cut in (a) and the result is displayed as a color plot **(d)** Distribution of $I_1/I_0$ as a function of angle ($\theta$) measured from the $+k_x$ direction. The distribution is obtained by radially integrating the surface state contours in (a) and (b) over a 'k' space window of width ~ 0.013 $A^{-1}$. **(e)** Calculated angular distribution of $I_1/I_0$ for P-polarized pump at different values of the LAPE parameter ($\alpha$) and the Floquet parameter ($\beta$). Red trace: $\alpha = 0$ and $\beta = 0.5$. Green trace: $\alpha = 1.38$ and $\beta = 0$. Blue trace: $\alpha = 1.38$ and $\beta = 0.5$.

14

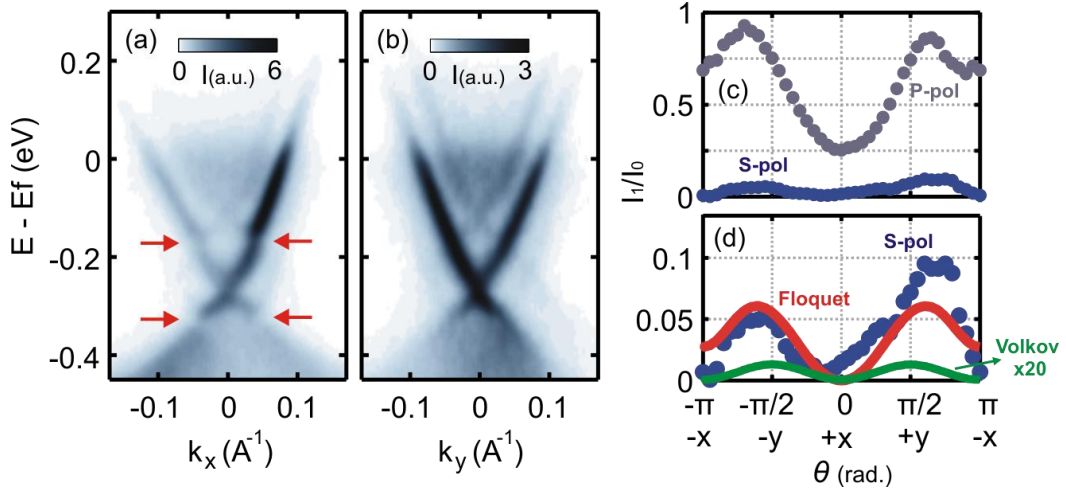

**Figure 4.** Tr-ARPES spectra at $t = 0$ for S-polarized pump. **(a)** Energy (E) relative to the Fermi level (Ef) vs momentum along the $k_x$ direction and **(b)** along the $k_y$ direction. Red arrows indicate the avoided crossing gaps. **(c)** Ratio of the first order side band intensity $I_1$ to the zeroth order intensity $I_0$ as a function of angle ($\theta$) measured from the $+k_x$ direction., for both P and S-polarized pump. **(d)** $I_1/I_0$ for S-polarized pump (blue trace) along with the calculated angular distribution of $I_1/I_0$ for S-polarized pump using: $\alpha = 0$ and $\beta = 0.5$ (red trace). Green trace represents the calculated $I_1/I_0$ for the case of $\alpha = 0.05$ and $\beta = 0$. These values correspond to Volkov states being generated by the in-plane electric field only (supplementary). Note: $I_1/I_0$ for the green trace has been multiplied by 20 for better visual representation on this axis scale.



\title{Supplementary information for: Selective scattering between Floquet-Bloch and Volkov states in a topological insulator}

\author{Fahad Mahmood, Ching-Kit Chan, Zhanybek Alpichshev, Dillon Gardner, Young Lee, Patrick A. Lee, Nuh Gedik}

\maketitle

\section{Theoretical details}

In this section, we provide details of calculations for the Tr-ARPES intensity from driven surface states of a topological insulator. We consider the intrinsic Floquet states attributed by the drive, the spin-probe effect on the photoemission matrix elements, and the influence of LAPE. These combined contributions lead to the theoretical results presented in the main text.

The effective Hamiltonian describing the undriven surface states of a 3D topological insulator is given by:
\begin{eqnarray}
H
&=&\sum_{k=k_x,k_y}
\begin{pmatrix}    c_{k,\uparrow}^\dagger & c_{k,\downarrow}^\dagger   \end{pmatrix}
\begin{pmatrix}    0 & \hbar v_f (-i k_x-k_y)  \\ \hbar v_f(i k_x-k_y) & 0 \end{pmatrix}
\begin{pmatrix}    c_{k,\uparrow} \\ c_{k,\downarrow}   \end{pmatrix},
\end{eqnarray}
where $c_{k,\sigma}^\dagger$ creates a bare electron with momentum $k$ and pseudospin $\sigma$. Diagonalizing the Hamiltonian would give the canonical linear dispersion of surface states, i.e. $\epsilon_\pm(\vec k) =\pm \hbar v_f |\vec k|$.  The laser drive is introduced through the Peierls' substitution: $ v_f \vec k\rightarrow \vec K(t)= v_f (\vec k+ e\vec A(t))$, with $\vec A(t)= A_0 g(t)\left(a_x\cos(\omega t),a_y\sin(\omega t)\right)$. $A_0$ is the peak pump field strength, $g(t)$ describes the Gaussian pump envelope and $0 \leq a_{x/y} \leq 1$ characterize its polarizations. We use $\beta= e v_f A_0/\omega$ as the dimensionless Floquet parameter in the main text.

The general idea of Floquet-Volkov transition has been studied by Park \cite{park14}. Using a scattering approach, it was estimated that the $n$-th Floquet sideband has an intensity (in our notation):
\begin{eqnarray}
I(\vec k,E) \approx J_n\left\{ \left| \lambda \beta   \left( \cos\theta_a \cos\theta +i \sin\theta_a \sin\theta \right) -\alpha \right| \right\}^2
\label{eq_park}
\end{eqnarray}
as a function of the momentum angle $\theta = \tan^{-1} (k_y/k_x)$ and the polarization angle $\theta_a = \tan^{-1} (a_y/a_x)$. $\lambda=\pm $ denotes the upper and lower bands, and $\alpha$ is a LAPE parameter to be defined later. However, we find that this formula does not fit well with the data because of approximations made (specifically, the components of $\vec A(t)$ perpendicular to $\vec k$ are neglected \cite{park14}), and more importantly, the missing of the spin-probe effect. Therefore, we introduce here an alternative method, based on the non-equilibrium Green's function \cite{freericks09}, that naturally incorporates the Floquet, the LAPE and the spin-probe effects and accurately describes our photoemission experiments.

\subsection{Tr-ARPES dynamics}

The pump-probe Tr-ARPES measures the photoelectron correlations \cite{freericks09}:
\begin{eqnarray}
I(\vec k,E)=\int_{t_i}^{t_f}dt \int_{0}^{t_f-t}d\tau s(t)s(t+\tau) \sum_{\sigma_1,\sigma_2,\sigma_f} M_k^*(\sigma_f,\sigma_1) M_k(\sigma_f,\sigma_2) 2~\text{Re}\left[ \langle c_{k,\sigma_1}^\dagger(t+\tau)c_{k,\sigma_2}(t)\rangle e^{-iE \tau/\hbar}\right],
\label{eq_TRAPRES}
\end{eqnarray}
caused by a probe field with a normalized Gaussian envelope $s(t)$, an initial time $t_i$ and a final time $t_f$. The expectation value is taken with respect to the wavefunction of the driven system. This expression describes a virtual process where an electron is photoexcited at time $t$ and then returns at time $t+\tau$. The matrix elements $M_k(\sigma',\sigma)$ correspond to transitions from spin $\sigma$ to $\sigma'$ and depend on the details of system-probe interactions. In the absence of spin-probe coupling, we have $M_k(\sigma',\sigma)\propto \delta_{\sigma,\sigma'}$ and the Tr-ARPES intensity becomes:
\begin{eqnarray}
I_0(\vec k,E) \propto \int_{t_i}^{t_f}dt \int_{0}^{t_f-t}d\tau s(t)s(t+\tau) \sum_{\sigma} 2~\text{Re}\left[ \langle c_{k,\sigma}^\dagger(t+\tau)c_{k,\sigma}(t)\rangle e^{-i E \tau/\hbar}\right].
\label{eq_TRAPRES_simple}
\end{eqnarray}
In the equilibrium limit, and for a uniform probe, this expression reduces to: $I_0(\vec k,E)\sim (t_f-t_i) \int_{0}^{\infty}d\tau  \sum_{\sigma} 2~\text{Re}[\langle c_{k,\sigma}^\dagger(\tau)c_{k,\sigma}(0)\rangle e^{-iE \tau/\hbar}]$, which is proportional to the standard lesser Green's function \cite{freericks09}.

To calculate Eq.~(\ref{eq_TRAPRES_simple}), we need the equation of motions for the two-time correlation functions. They are:
\begin{eqnarray}
\frac{d}{d \tau} \langle c_{k,\uparrow}^\dagger(t+\tau) c_{k,\uparrow/\downarrow}(t)\rangle &=& (- K_x(t+\tau)-i K_y(t+\tau))  \langle c_{k,\downarrow}^\dagger(t+\tau) c_{k,\uparrow/\downarrow}(t)\rangle, \nn \\
\frac{d}{d \tau} \langle c_{k,\downarrow}^\dagger(t+\tau) c_{k,\uparrow/\downarrow}(t) \rangle &=& ( K_x(t+\tau)-i K_y(t+\tau))  \langle c_{k,\uparrow}^\dagger(t+\tau) c_{k,\uparrow/\downarrow}(t)\rangle.
\label{eq_timecorrelation}
\end{eqnarray}
The initial condition at $\tau=0$ is determined by the equal-time correlations $\langle c_{k,\sigma}^\dagger(t) c_{k,\sigma'}(t) \rangle$. \textbf{Generally speaking, the initial condition $\langle c_{k,\sigma}^\dagger(t) c_{k,\sigma'}(t) \rangle$ depends on $t$ due to the non-equilibrium nature of the problem. However, in the main text, we are mostly interested in the cases of  $|\vec k|\leq k_F$, where both the lower and upper band of the undriven Dirac cone are filled so that no vertical transition is allowed by the drive. In this regime, we have $\langle c_{k,\sigma}^\dagger(t) c_{k,\sigma'}(t) \rangle = \langle c_{k,\sigma}^\dagger(0) c_{k,\sigma'}(0) \rangle = \delta_{\sigma, \sigma'}$. On the other hand, for $|\vec k|> k_F$, $\langle c_{k,\sigma}^\dagger(t) c_{k,\sigma'}(t) \rangle$ becomes t-dependent and can be computed by solving its equation of motion using the driven Hamiltonian.}

\textbf{
Since the relaxation time scale is much longer than that of the drive in our experiment, dissipative effects are negligible. However, when dissipation becomes important, we have to augment Eq.~(\ref{eq_timecorrelation}) with relaxation terms based on the master equation formalism \cite{dehghani14} or the kinetic approach \cite{seetharam15}.
}

\subsection{Spin-probe effect}

When the probe couples to the spin, one has to compute the matrix elements $M_k(f,i) = \langle f | \vec{P}_k \cdot \vec{A}_{\text{probe}}(t) | i\rangle$ \cite{wang11}. In accordance with our experimental setup, we consider a p-polarized probe such that $\vec{A}_{\text{probe}}=(A_x,0,A_z)$. In general, $\vec{P}_k$ is some generalized momentum and depends on the system details and spin-orbit couplings. However, by considering the mirror reflection symmetry and the three-fold rotational symmetry \cite{wang11}, one can simplify the matrix elements to (in the basis of spin up and down):
\begin{eqnarray}
M_k=
i\begin{pmatrix}    b_k A_z & \frac{a_k A_x}{2}  \\  -\frac{a_k A_x}{2} & b_k A_z \end{pmatrix},
\label{eq_matrixelement}
\end{eqnarray}
where $a_k$ and $b_k$ are real coefficients. Using these matrix elements, Eq.~(\ref{eq_TRAPRES}) becomes:
\begin{eqnarray}
I(\vec k,E)\propto  I_0(\vec k,E)+  c_{SP} \int_{t_i}^{t_f}dt \int_{0}^{t_f-t}d\tau s(t)s(t+\tau)  2~\text{Re}\left[ e^{-iE \tau/\hbar}   \langle -i c_{k,\uparrow}^\dagger(t+\tau)c_{k,\downarrow}(t) + i  c_{k,\downarrow}^\dagger(t+\tau)c_{k,\uparrow}(t)\rangle \right],
\label{eq_TRAPRES_spinprobe}
\end{eqnarray}
where the parameter $c_{SP}=[ia_k b_k A_z^* A_x +\text{c.c.}]/[a_k^2|A_x|^2/2 + 2b_k^2 |A_z|^2]$ characterizes the spin-probe effect. Note that the spin-probe effect only couples to the $S_y$ component in this setting of a p-polarized probe. \textbf{Because of the spin-momentum locking, we have $S_y \sim \cos \theta$ and thus the spin-probe effect is most prominent at $\theta = 0, \pi$, i.e. along the $k_x$ direction. For a probe with a general polarization, the $S_x$ and $S_z$ components can also be involved ($S_x$ couples to $A_{y,z}$ and $S_z$ couples to $A_{x,y}$).}

We can extract the single coefficient $c_{SP}$ from the pre-time zero data. In the equilibrium limit, for a p-polarized probe, we recover the static result: $I - I_0 \propto c_{SP} S_y$ \cite{wang11}.

\subsection{LAPE}

The Laser-Assisted Photoemission Effect (LAPE) is caused by the interference between the pump and the photoexcited electrons. The corresponding Hamiltonian is $H_{\text{LAPE}} = \hbar e \vec{v}_0 \cdot \vec{A}_{\text{pump}}(t)$, where $\vec{v}_0$ is the free photoelectron velocity. Consequentially, the two-time correlation function in the Tr-ARPES intensity [Eq.~(\ref{eq_TRAPRES})] picks up an additional phase:
\begin{eqnarray}
e^{i \Theta(\alpha_k,t,\tau)} &=&  e^{-i \int_t^{t+\tau}  dt' H_{\text{LAPE}}(t')/\hbar}    \nn \\
 &=&\exp \{-i (\alpha_{k_x}+\alpha_{k_z}) [\sin \omega(t+\tau) -\sin \omega t     ] +i \alpha_{k_y} [\cos \omega(t+\tau) -\cos \omega t     ]\},
\label{eq_phasefactor_1}
\end{eqnarray}
where $\alpha_{k_i} = e v_{0,i} A_{\text{pump},i}/\omega$. Note that the pump field can have a $z$-component contribution to the LAPE. In fact, for the p-polarized pump experiment, we have $\alpha_{k_z} \gg \alpha_{k_x/k_y}$, since the pump field has a larger $z$-component and $|v_{0,z}|\sim v_f \gg |v_{0,x/y}|$. Thus, we set $\alpha_{k_z}=\alpha$ and $\alpha_{k_x/k_y}=0$ in the main text. On the other hand, in the s-polarized pump case, all $\alpha_k$ are negligible.

Therefore, \textbf{for a p-polarized probe,} the Tr-ARPES intensity, including the spin-probe ($c_{SP}$) and the LAPE ($\alpha_k$) effects, is given by:
\begin{eqnarray}
I(\vec k,E)\propto \int_{t_i}^{t_f}dt \int_{0}^{t_f-t}d\tau s(t)s(t+\tau)  2~\text{Re}\left\{ e^{-iE \tau/\hbar}  e^{i \Theta(\alpha_{k},t,\tau)} \left[ S_0(t,\tau)  +c_{SP} S_y(t,\tau) \right] \right\},
\label{eq_TRAPRES_spinprobe_LAPE}
\end{eqnarray}
where $S_0(t,\tau) = \sum_\sigma \langle c_{k,\sigma}^\dagger(t+\tau)c_{k,\sigma}(t)\rangle $ and $S_y(t,\tau) =  \langle -i c_{k,\uparrow}^\dagger(t+\tau)c_{k,\downarrow}(t)   + i c_{k,\downarrow}^\dagger(t+\tau)c_{k,\uparrow}(t) \rangle $ can be obtained by solving Eq.~(\ref{eq_timecorrelation}). The theoretical curves in Figure 3e and 4d in the main text are calculated from this expression.

\subsection{Exactly solvable case: parity of Floquet-LAPE states}

Analytical solutions are available for special cases where $\vec k \parallel \vec A(t)$. Here we particularly study the situation of (1) $\pm k_x$ for a p-polarized pump, and (2) $\pm k_y$ for a s-polarized pump, in order to understand the parity of the Tr-ARPES signals with and without the LAPE effect. We use uniform pump and probe to simplify the notations.

\subsubsection{Without LAPE}

For the case (1), the driven Hamiltonian $H(t) = \hbar v_f \left(k_x +e A_0 \cos\omega t   \right) \sigma_y$. The two-time correlation functions can be solved from Eq.~(\ref{eq_timecorrelation}) and we find:
\begin{eqnarray}
S_0(t,\tau) &=& 2 \cos \left\{ v_f k_x\tau + \beta \left[\sin \omega(t+\tau) -\sin \omega t \right]  \right\} \nn \\
&=& \sum_{n_1,n_2} e^{i (v_f k_x +n_1 \omega)\tau}e^{i (n_1+n_2) \omega t} J_{n_1} \left(\beta \right) J_{n_2} \left(-\beta \right) +c.c. , \nn \\
S_y(t,\tau) &=& 2 i \sin \left\{ v_f k_x \tau + \beta \left[\sin \omega(t+\tau) -\sin \omega t \right]  \right\} \nn \\
&=&  \sum_{n_1,n_2} e^{i (v_f k_x +n_1 \omega)\tau}e^{i (n_1+n_2) \omega t} J_{n_1} \left(\beta \right) J_{n_2} \left(-\beta \right) -c.c. ,
\label{eq_timecorrelation_2}
\end{eqnarray}
Using these expressions, we can perform the double-time integral in Eq.~(\ref{eq_TRAPRES_spinprobe_LAPE}) for $|k_x|\leq k_F$. The $\tau$ integral leads to Tr-ARPES peaks, while the $t$ integral averages over different Fourier modes, leaving only the $n_1=-n_2$ components. The Floquet result for a p-polarized pump is:
\begin{eqnarray}
I (k_x,E) \propto T \sum_n \left\{ (1+c_{SP})\delta_{E,\hbar v_f k_x+n \hbar\omega} +   (1-c_{SP})\delta_{E,-\hbar v_f k_x+n \hbar \omega} \right\} J_{n} \left(\beta \right)^2,
\label{eq_TRAPRES_noLAPE}
\end{eqnarray}
where $T=t_f-t_i$ is the probe duration. The $n$-th intrinsic Floquet peak (when $c_{SP}=0$) has an intensity of $J_{n} \left(\beta\right)^2$ and is symmetric about $\pm k_x$. The spin-probe effect breaks this parity. Similarly, for the case (2) of a s-polarized pump, we find
\begin{eqnarray}
I (k_y,E) \propto T \sum_n \left\{ \delta_{E,\hbar v_f k_y+n \hbar\omega} +   \delta_{E,-\hbar v_f k_y+n \hbar \omega} \right\} J_{n} \left(\beta \right)^2,
\label{eq_TRAPRES_noLAPE2}
\end{eqnarray}
where the spin-probe effect does not enter due to the absence of the $S_y$ component along $k_y$.

\subsubsection{With LAPE}

The situation becomes very different in the presence of the LAPE. We first consider the case (1) of $\pm k_x$ for p-polarization with a constant LAPE strength $\alpha_{k_z}=\alpha$. The LAPE effect gives rise to a phase [Eq.~(\ref{eq_phasefactor_1})]:
\begin{eqnarray}
e^{i \Theta(\alpha,t,\tau)}&=&e^{-i \alpha [\sin \omega(t+\tau) -\sin \omega t     ]}   \nn \\
&=& \sum_{m_1,m_2} e^{i m_1 \omega \tau}e^{i (m_1+m_2) \omega t} J_{m_1} \left(-\alpha\right) J_{m_2} \left(\alpha \right).
\label{eq_phasefactor_2}
\end{eqnarray}
Inserting Eqs.~(\ref{eq_timecorrelation_2}) and~(\ref{eq_phasefactor_2}) into Eq.~(\ref{eq_TRAPRES_spinprobe_LAPE}), the Tr-ARPES intensity for p-polarization becomes:
\begin{eqnarray}
I (k_x,E) &\propto & T (1+c_{SP}) \Big[\sum_{n_1,m_1}  \delta_{E,\hbar v_f k_x+(n_1+m_1) \hbar\omega}\ J_{n_1} \left( \beta\right) J_{m_1} \left(-\alpha\right) \Big]^2 \nn \\
&& \ \ \ \ \ \ +\ T (1-c_{SP}) \Big[\sum_{n_1,m_1}   \delta_{E,-\hbar v_f k_x-(n_1-m_1) \hbar \omega}\ J_{n_1} \left( \beta\right) J_{m_1} \left(-\alpha\right) \Big]^2 \nn \\
&\propto & T \sum_{n}  \Big[ (1+c_{SP}) \delta_{E,\hbar v_f k_x+ n \hbar \omega}\ J_{n} \left( \beta-\alpha\right)^2 + (1-c_{SP}) \delta_{E,-\hbar v_f k_x + n \hbar \omega}\  J_{n} \left( \beta+\alpha\right)^2 \Big]  .
\label{eq_TRAPRES_LAPE}
\end{eqnarray}
In comparison with Eq.~(\ref{eq_TRAPRES_noLAPE}), these equations demonstrate how the LAPE effect interferes with the intrinsic Floquet peaks. For example, for the zeroth order peak at $E=\hbar v_f k_x$, we can have contributions from different Fourier pairs of Floquet ($n_1$) and LAPE ($m_1$) modes that satisfy $n_1+m_1=0$ as shown in the first line of equations.

We can extract the peak intensities for fixed $E$ and $k_x$. The result depends on the parity of $k_x$. At the crossing points, i.e. when $2 v_f k_x/\omega $ becomes  an integer, we have
\begin{eqnarray}
I (k_x>0,E=\hbar v_f |k_x|+n \hbar\omega)&\propto & (1+c_{SP})J_{n} \left( \beta-\alpha\right)^2 + (1-c_{SP})J_{n+2 v_f |k_x|/\omega} \left( \beta+\alpha\right)^2,  \nn \\
I (k_x<0,E=\hbar v_f |k_x|+n \hbar\omega)&\propto & (1-c_{SP})J_{n} \left( \beta+\alpha\right)^2 + (1+c_{SP})J_{n+2 v_f |k_x|/\omega} \left( \beta-\alpha\right)^2.
\label{eq_TRAPRES_LAPE_nth}
\end{eqnarray}
The first and second terms in each equation describe the Floquet sideband contributions coming from the upper and lower branches, respectively. In this case of a p-polarized pump, even in the absence of the spin-probe effect, these two expressions are in general different and the $\pm k_x$ signals are asymmetric in each Floquet band, in accordance with the data presented in Figure 3 in the main text.

We can work out the case (2) of $\pm k_y$ for a s-polarized pump in the same way. The resultant Tr-ARPES intensity is:
\begin{eqnarray}
I (k_y,E) &\propto & T \sum_{n}  \Big[ \delta_{E,\hbar v_f k_y+ n \hbar \omega}\ J_{n} \left( \beta-\alpha_{k_y}\right)^2 + \delta_{E,-\hbar v_f k_y + n \hbar \omega}\  J_{n} \left( \beta+\alpha_{k_y}\right)^2 \Big]  .
\label{eq_TRAPRES_LAPEs}
\end{eqnarray}
Again, the spin-probe effect does not appear because of the vanishing of the $S_y$ component along $k_y$. We have included a small but finite LAPE parameter $\alpha_{k_y} \propto k_y$. Different from the p-polarization situation, $\alpha_{k_y}$ changes sign between $+k_y$ and $-k_y$ here. Thus, the Floquet sideband weights remain symmetric between $+ k_y$ and $- k_y$. Based on this result, the slight asymmetry at $\pm k_y$ observed in Figure 3d in the main text are not caused by the LAPE effect.

We note that in the absence of the spin-probe effect and when $\vec k \parallel \vec A(t)$, our results [Eq.~(\ref{eq_TRAPRES_LAPE},~\ref{eq_TRAPRES_LAPEs})] agree with Park's expression [Eq.~(\ref{eq_park})]. In general, an analytical solution is not available and we have to compute the Tr-ARPES intensity numerically from Eq.~(\ref{eq_TRAPRES_spinprobe_LAPE}).

\section{Experimental Parameters}

In this section we provide estimates of various experimental parameters. As discussed in the main text, the two dimensionless parameters relevant to this work are $\alpha$ and $\beta$. $\alpha = ev_0A_0/\omega$ characterizes the interaction strength between light and the final states of photoemission while $\beta = e v_f A_i/\omega$ characterizes the strength of the Floquet interaction. Here $A_{0,i} = E_{0,i}/\omega$ where $E_{0,i}$ is the electric field amplitude along a particular electron velocity direction. For Floquet states, the relevant velocity is the Fermi velocity for the surface state electrons. Since this velocity is purely in-plane, the relevant electric field ($E_i$) is the one parallel to the sample surface. Taking $E_i = 3.3\times10^7 $  V/m corresponding to a measured pump power of $11.5$ mW (see detailed explanation of estimate in supplementary section of Ref.\cite{wang13}) and $v_f = 5\times10^5$ m/s, we obtain $\beta = 0.42 $. \vspace{\baselineskip}

For the LAPE effect we need to determine the electron velocity in the final state of photoemission. As the in-plane momentum is conserved in the photoemission process and given that the final state is free electron-like, we determine the in-plane velocity, $v_{\parallel} =  5.79\times10^4$ m/s for momentum, $k = 0.05$ A$^{-1}$. By conserving energy, this gives the out-of-plane electron velocity, $v_z = 4.55\times10^5$ m/s . Note that $v_z \gg v_{\parallel}$. Thus, the relevant velocity for the LAPE effect is $v_0=v_z$ and the relevant electric field ($E_0$) is the out-of-plane component of the electric field outside the sample surface. Using Fresnel equations, we obtain $E_0 = 11.6\times10^7$ V/m and thus $\alpha \sim 1.36-1.4$. \vspace{\baselineskip}

Note the values of $\alpha = 1.38$ and $\beta = 0.5$ used in the main text are determined by fitting the observed angular dependence of the sideband intensities in the Tr-ARPES spectrum to the theoretically calculated intensities in section I. We also used $c_{SP}=0.96$ and $0.7$ for the p- and s-polarized pumps, respectively, to account for the spin-probe effect. Given the large uncertainty in determining the exact electric field at the sample surface, these values are consistent with the values calculated in this section. The full widths at half maximum for the pump and the probe are 250~fs and 100~fs, respectively.

\section{Determination of hybridization gap}

The avoided crossing gaps in Fig. 2c of the main text are determined by taking an Energy Distribution Curve (EDC) at $k_y = 0.035$ A$^{-1}$ through the Tr-ARPES spectra obtained at $t = 0$ for each pump power. This value of $k_y$ corresponds to where the avoided crossing gap is observed. The EDC for the Tr-ARPES spectra in Fig. 2b of the main text is shown in Fig. S1. Each EDC is fitted with multi-peak Lorentz functions. The choice of the peak functions does not affect the obtained gap size in any significant way since we are only interested in the distance between the peaks. The avoided crossing gap is then given by the separation of the two peaks around energy, $E-E_f = -0.2$ eV (green peaks in Fig. S1). The resulting gap size for a pump power of $11.5$ mW is then $2\Delta = 68$ meV  which is in good agreement with the experimental parameters stated in section II as $2\Delta = \beta \omega$ with $\omega = 160$ meV and the $\beta = 0.42$. \vspace{\baselineskip}

This procedure is also repeated for the case of S-polarized pump and the resulting gap ($2\Delta$) as a function of pump power ($P$) is plotted in Fig. S2 on a log-log plot. Similar to the case of P-polarized pump, $2\Delta$ scales as the square root of the pump power i.e. $2\Delta \propto \sqrt{P}$ in agreement with Floquet-Bloch theory on Dirac systems
\cite{fregoso13}.

\renewcommand{\figurename}{Fig. S}
\setlength{\belowcaptionskip}{15pt}

\begin{figure}[h]
\caption{Energy Distribution Curve (EDC) at $k_y = 0.035$ A$^{-1}$ through the Tr-ARPES spectra in Fig. 2b of the main text. Blue dots indicate the raw data. The red line is a multi-peak Lorentz function best fit to the data. Green peaks correspond to the peaks from which the avoided crossing gap is determined}
\centering
\includegraphics[width=0.5\textwidth]{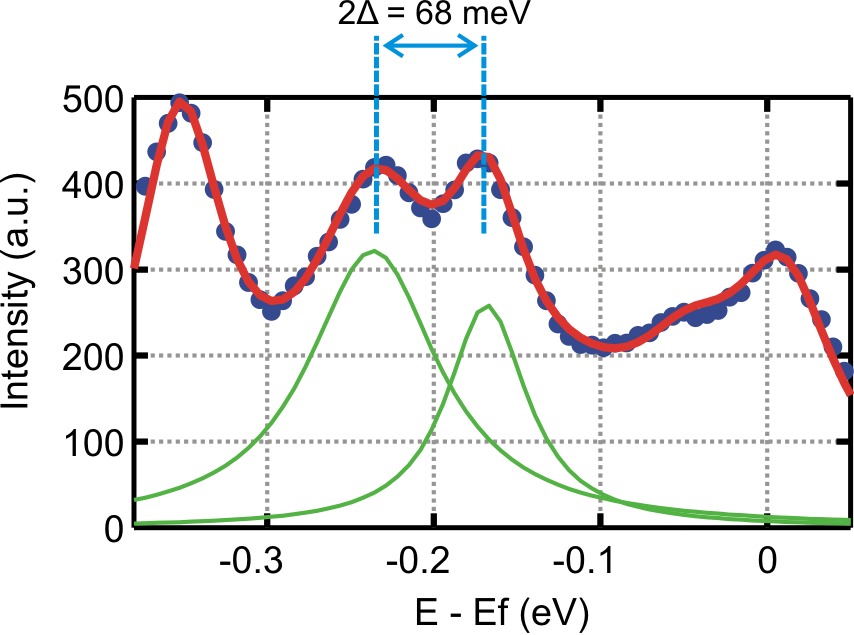}
\end{figure}

\begin{figure}[h]
\caption{Avoided crossing gap ($2\Delta$) as a function of incident pump power ($P$) on a log-log plot for S-polarized pump. Error bars represent the 95\% confidence interval (2 s.d.) in extracting the gap from the fitting parameters. Power laws ($2\Delta \propto P^\eta$) with $\eta = 0.5$ (orange trace) and $\eta= 1$ (blue trace) are plotted as well to determine the analytical behavior of $2\Delta$ with $P$.}
\centering
\includegraphics[width=0.4\textwidth]{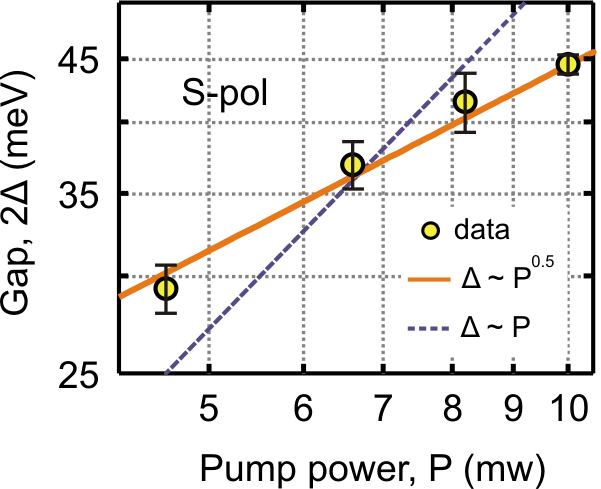}
\end{figure}


\begin{thebibliography}{99}
\bibitem{freericks09} J. K. Freericks, H. R. Krishnamurthy, and Th. Pruschke, Phys. Rev. Lett. \textbf{102}, 136401 (2009).
\bibitem{dehghani14} H. Dehghani, T. Oka, and A. Mitra, Phys. Rev. B \textbf{90}, 195429 (2014).
\bibitem{seetharam15} K. I. Seetharam, C.-E. Bardyn, N. H. Lindner, M. S. Rudner, and G. Refael, arXiv:1502.02664 (2015).
\bibitem{wang11} Y. H. Wang, D. Hsieh, D. Pilon, L. Fu, D. R. Gardner, Y. S. Lee, and N. Gedik, Phys. Rev. Lett. \textbf{107}, 207602 (2011).
\bibitem{park14} S. T. Park, Phys. Rev. A \textbf{90}, 013420 (2014).
\bibitem{wang13} Y. H. Wang, H. Steinberg, P. Jarillo-Herrero, N. Gedik, Science \textbf{342}, 453-457 (2013).
\bibitem{fregoso13} B. M. Fregoso, Y. H. Wang, N. Gedik and V. Galitski, Phys. Rev. B \textbf{88}, 155129 (2013).


\end{thebibliography}
\end{document}